# A Review of Empirical Applications on Food Waste Prevention & Management

Ahmed Fadhil
University of Trento
Trento, Italy
fadhil@fbk.eu

*Abstract*—Food waste has a significant detrimental economic, environmental and social impact. Recent efforts in HCI research have examined ways of influencing surplus food waste management. In this paper, we conduct a research survey to investigate and compare the effectiveness of existing approaches in food waste management throughout its lifecycle from agricultural production, post-harvest handling and storage, processing, distribution and consumption. The objectives of the survey are 1) to identify methods in food waste management, 2) their area of focus, 3) the ICT techniques they apply, 4) and the food waste lifecycle they target. In addition, we analyse if 5) they provide an open access API for food waste data analysis. Based on the literature analysis, we then highlight their pros and cons with respect to applications in food waste management. The implications of this research could present a new opportunity for interested stack-holders and future technologies to play a key role in reducing domestic and national food waste.

*Keywords*-Food waste management, gamification, waste phases, ICT for sustainability.

## I. INTRODUCTION

About 1/3 of the food produced for human consumption is lost or wasted; approximately 1.3 billion tonnes per year [1]. Annually, 88 million tonnes of food are wasted in Europe, with an associated cost estimated at e143 billion. Developed and developing countries dissipate roughly the same amount of food, respectively 670 and 630 million tonnes [1]. Moreover, per capita waste by consumers is between 95-115 kg a year in Europe and North America, while its around only 6-11 kg a year in sub-Saharan Africa and south-eastern Asia. Over 40% of food losses occurs in industrialised countries at retail and consumer level, whereas roughly similar amounts in developing countries occurs after harvesting and during processing [1].

Food waste happens at households, supermarkets, and restaurants. For instance, domestic food waste occurs during food purchasing and once food enters the home [2]. To decrease this burden, countries have set regulation to force food suppliers and consumers to take preventive actions. For example, France has introduced taxation on food wasted at the supermarkets; Italy has offered tax breaks on supermar-kets donating surplus food to charities. Nonetheless, edible food is still often thrown away due to lack of knowledge on food planning or donation regulation. Technology can promote changes and help make a difference in the habit of food waste. Existing literature provides insight into a range of applications intended for food waste management and prevention. These studies provide insights about waste generated at different phases, namely production (e.g., farming and manufacturing), distribution (e.g., food marketing, such as the selling process in a supermarket), and consumption (e.g., food consumption in restaurants and homes). Moreover, these applications use one or more tech-niques to create the necessary infrastructure for the services with which they contribute to food waste management.

This paper contributes to the understanding of food waste management by reviewing the existing body of empirical applications on the topic. That said, we will discuss the focus area, technology and phases for the analysed applications. Moreover, we will highlight if APIs or data analyses are provided by these applications for third parties. The focus area refers to the way these applications approach food waste. For example, the application can provide food dona-tion, food sharing, coupon promotion about food near their expiry date, or even food that is used in composting. The technology focus refers to the technological approach used. This paper covers geolocation, gamification, crowdsourcing and quantified-self techniques in food waste management applications (see Table I).

Table I
APPROACHES IN FOOD WASTE MANAGEMENT.

| Approaches | Description |
|---|---|
| Geolocation | The wireless detection of the physical location of a remote device [3]. |
| Gamification | The concept of applying game mechanics and game design techniques to engage and motivate people to achieve their goals [4]. |
| Crowdsourcing | The practice of obtaining services, or content by contri-butions from online communities rather than traditional employees or suppliers [5]. |
| Quantified-Self | The incorporation of technology into data acquisition on aspects of a person's daily life in terms of inputs, states, and performance. [6]. |

The different phases include production, distribution, and consumption as given above. Based on the literature findings, we will propose an approach on how to improve such applications to be more engaging and motivating for the user. This could provide a new direction on how to deal with food waste with a deeper level of understanding on how current trends deal with food waste and what could be done to improve the academic and industrial research in the domain.

## II. REASONS FOR FOOD WASTE

Along the food supply chain, food waste occurs during production, distribution and consumption. For example, in the US of all food waste, production waste comprises 20%, processing 1%, distribution 19%, and 60% of food waste generated by consumers [7]. Moreover, less than 1/3 of food waste is recovered by composting 25% and donations 3%, and over 72% are landfilled [7].

In general, vegetables, animal products, and commodities are the most contributors to the generated waste among all food supply chains (FSC) segment. To illustrate, losses among agricultural production are due to mechanical damage or spillage during harvest operations (e.g. thrashing or fruit picking). However, wastes during post-harvest handling/stor-age is due to spillage and degradation during transportation between farm and distribution. Whereas, during consump-tion segment, the waste occurs at the household level.

Moreover, large packaging, poor quality of purchased gro-ceries, family income, cooking too much due to lack of experience are all contributors to food waste at household. The age of the consumer is also a significant factor in the food waste. For example for the people aged over 65, many studies come to the conclusion that they waste significantly less food than the rest of the population [8].

To reduce this burden, approaches, such as donation have grown significantly. Nearly half of the food collected in Europe comes from the European programme of food aid for the most deprived (FEAD) 33%, 22% from the food industry (manufacturers), 17% from the retail sector and 14% from individuals through national and local collections [7].

## III. FOOD WASTE MANAGEMENT APPLICATIONS

This section examines a range of applications in food waste management. We discuss techniques applied, phases and technological support for each application in supporting food waste management.

Several studies have focused on food waste reduction on consumer level; for instance, LELOCA [9] is an application to reduce the amount of fresh food waste in restaurants. Leloca provides a geolocalized dining coupons by tagging them on the map. In this way, restaurants decrease their surplus food waste. Researchers from University of Twente have developed CHEETAH [10] to reduce crops waste in West Africa. The application exchanges traffic data between food transporters, growers and traders. This would minimise food spoilage risk due to late delivery. Therefore, Cheetah would help transporters, growers and traders to exchange best route information. In a different study BREADING [11] was developed to minimise bread leftovers at bakeries. The application allows bakers to donate their leftover bread and manage its collection process. The donation are geolocalized for nearby collectors to get noticed. In a similar way, BRINGTHEFOOD [12] manages all type of donations and at all phases.

RATATOUILLE [13] was developed to work at personal con-sumption level and allow people to donate food from their fridge. The application uses geolocation to display nearest fridge and gives an expiry date for each food donation. In addition, Ratatouille is used by hostels and students who are accustomed to share leftovers.

Not to mention that food waste is strongly correlated with user's behaviour. Maintaining specific behaviour over time is a hard task [14], [15], [16]. Blevis et al. (2007) [17] discusses the effect of motivational techniques, such as gamification on maintaining a behaviour. Blevis states that research must consider engagement integration into tech-nology to stimulate behaviour change process. Many food waste applications have gained initial interests by users to stimulate behavioural changes, but they failed to sustain it. Few approaches have applied gamification technique in food waste. For instance, MINTSCRAPS [19], a platform to empower restaurants monitor waste generation patterns, by tracking what food type is wasted, its quantity and at which mealtime. The user has to insert their leftovers to get precise waste data. The platform uses gamification to increase user's engagement and awareness in waste management practices. The application provides insights for users to self-manage their leftovers. This approach encourages self-planing to better manage the waste generated. Even with healthier food access, to achieve behaviour change, games should educate nutritional attributes of food, help build knowledge base of better or worse food choice, and help develop skills to inter-pret nutrition information. For example, FITTER CRITTERS [20] is a game to teach 8-12 years old kids healthier eating habits. The player is responsible for maintaining the health of a virtual pet. To do that, they need to shop for critter's food, cook for it, feed and exercise it.

A work by Unilever[1] developed WISE UP ON WASTE [21] application for professional kitchens to track their waste generation and potential cost savings. In the context of meal planning, MENUS4MUMS[22] website provides online meal plans, recipes and corresponding shopping list for families. The uniqueness of this service is including products with offers in the menu. This website promotes self-management of food leftovers by ensuring all ingredients bought at the start of the week are used up in the meal.

Some applications try to cut food waste by providing people with recipes based on leftovers in the kitchen and advices on how to reduce it. For instance, LOVE FOOD HATE WASTE [23] and SMART SAVING [24] both provide users with a search functionality based on leftover ingredients. Such applications educate individuals to self-track their leftovers and turn them into a meal using the provided recipes. However, the database in these applications is limited to a fixed dataset. On the other hand, other applications use crowdsourcing to get updated data with different recipes. To illustrate, food recipes and photos are shared with new recipes which is vital for people to use the application. Similarly, S-CAMBIA CIBO [25], SHARECITY [26], BRINGTHEFOOD [12], FOOD COWBOY [27]; all rely on the crowd to share food or products approaching expiry date with others. In addition, these applications are useful for charities to collect food and provide and redistribute it to the deprived. SPOILER ALERT [28] an application developed by MIT students works in the same way, it also targets organic waste from which farms, supply chains and nonprofits can create value. A Swedish initiative at Malmborgs Tuna supermarket have developed RESOURCEFUL CHEF [29] where chefs transform discoloured fruits, wrinkly veggies and goods approaching their expiration date into meals for customers to purchase from the store. The store have created promotional healthy meals for customers and reduce food waste in supermarkets. SHARE YOUR MEAL[30] is another tool that allows users to see what people are cooking in their area to order a portion or two of their meal. This tool contributes to food waste reduction by letting people offer their additional por-tions that would have otherwise gone on waste. LEFTOVER SWAP [31] is an application that lets you post a photo of any leftovers you have. Others in the area can come and collect the leftovers. MADEFOOD [32] provides takeaways service of frozen meals. These meals are homemade and fresh. The application provides specific meal portions, hence reduces over ordering. FOODSTAR[29] is a platform that alerts shoppers of discounts at their local grocery store on imperfect or ripe produce. The platform reduces food waste by selling them through the platform. CROPSMOBSTER [33] is a platform, where farmers can sell off produce at a cheap price as it can't be sold to shops. For example, produce could be too near its use-by date, or visually damaged, but still edible. This largely reduces the amount of food waste. FOODSHARING [34] is a platform that allows individuals, traders and manufacturers to offer or collect leftover or unwanted food for free. Users can meet their people and share ideas through the platform. However, the way to look for available food is inefficient.

---
[1] https://www.unilever.com/

In a different context, HELLO COMPOST [35] an application to incentives families collect their food waste by rewarding them with credits for fresh produce. The service provides collection bags for low-income families to fill with food waste. Then, the families deliver the waste to project centre and earn credits based on the weight of the bag. Earned credits are redeemed for locally grown produce at the centre. On the other hand, the project centre turns the waste into compost which they sell to help fund the project. The data collection is used to visualise the positive impact residents are making in their community. Targeting low-income area, where people have low budgets for food made receiving food for free particularly appealing. This application used self-quantification to educate families manage their waste. Moreover, credits was used as points to get the free food which are the rewards. Finally, Hello Compost helped to improve the diet of some community members by providing cheap access to fresh produce.

Other applications in food waste include MYFOODY [36], LAST MINUTE SOTTO CASA [37], IFOODSHARE [38], SENZASPRECO [39], and FOOD CLOUD [40] which we discuss in next sections.

IV. ANALYSIS OF EXISTING APPLICATIONS

The survey listed food waste management applications based on their focus area, techniques, phases, data analysis, status, medium and location. More specifically, we anal-ysed techniques (geolocation, gamification, crowdsourcing, quantified-self), phases (production, distribution, and consumption), and the focus area for the applications (donation, sharing, promotional coupons, composting, etc). In addition, we have investigated if any of the applications collects data about food waste generation or provides API's. Tables II, III, and IV list each application and their correlation with various features based on the above set measures. For example, BRINGTHEFOOD geolocalizes food donations on the map. It also relies on the crowds to contribute in publishing their donations. This application manages donations at all phases. For instance, a donation could come from the farm, supermarket, or home. The application provides an API for food donation and waste reduction.

A number of conclusion can be drafted from the listed tables. Generally, out of 32 applications, gamification was found only in 4 applications [19], [18], [20], [35]. Moreover, most of the applications target surplus food at consumption phase. Few applications cover all phases, most focus on consumption phase and few have considered collecting data about food waste generation. In what follows, we further discuss the result of the survey analysis by providing an overview for each feature presented in the application. We then discuss the effectiveness of the applications in terms of user adoption, awareness and knowledge, needs, engagement and attitude and behaviour change.

V. FINDINGS

To carryout a methodological evaluation of the survey findings we investigate whether the applications provide measure for user adoption, awareness and knowledge, needs, engagement and attitude and behaviour change in their approaches to fight food waste. We analyse if these applications motivate users be more responsible and if they use persuasive techniques to promote their waste behaviour. Moreover, whether they focus on preventing waste rather than managing it. Finally, if the applications focus on user engagement and create a fun environment for the user while interacting with the application.

Table II
THE APPLICATIONS AND THEIR FOCUS AREA.

| Focus Area | Applications |
|---|---|
| Food Waste | Leloca, Cheetah, Breading, BTF, Ratatouille, MintScraps, WiseUpOnWaste, LoveFoodHateWaste, SmartSaving, S-Cambia Cibo, ShareCity, Food Cowboy, Resourceful Chef, ShareYour Meal, Leftover Swap, MadeFood, FoodStar, CropsMobster, Foodsharing, Hello Compost, MyFoody, LastMinuteSottoCasa, IFoodShare, SenzaSpreco, Gojee. |
| Promotional Discounts | Leloca, Menus4Mums, Fitter,Critters, Resourceful Chef, FoodStar, MyFoody, SenzaSpreco. |
| Food Donation | Breading, BTF, S-Cambia Cibo, ShareCity, Food Cowboy, Spoiler Alert, Leftover Swap, MyFoody, Food Cloud. |
| Surplus Food | Breading, S-Cambia Cibo, ShareCity, Food Cowboy. |
| Meal Planing | Menus4Mums. |
| Energy Savings | Polar Bear. |
| Charity | Spoiler Alert, FameZero, IFoodShare, SenzaSpreco. |
| Organic Waste | Kroger Co., Spoiler Alert. |
| Food Sharing | ShareYour Meal, Leftover Swap, MadeFood, CropsMobster, CropsMobster, Hello Compost. |
| Healthy Diet | Hello Compost. |

Table III
THE APPLICATIONS AND THEIR TECHNIQUES.

| Techniques | Applications |
|---|---|
| Geolocation | Leloca, Cheetah, Breading, BTF, Ratatouille, ShareYourMeal, Leftover Swap, MadeFood, FoodStar, Foodsharing. |
| Crowdsourcing | BTF, LoveFoodHateWaste, SmartSaving, S-Cambia Cibo, ShareCity, Food Cowboy, Spoiler Alert, FoodCloud, Gojee. |
| Gamification | PolarBear, MintScraps, MintScraps, Hello Compost. |
| Quantified-self | PolarBear, MintScraps, WiseUpOnWaste, Menus4Mums, Fitter Critters, LoveFoodHateWaste, SmartSaving, ShareYour Meal, Hello Compost, The Kroger Co. |

Table IV
THE APPLICATIONS AND THEIR FOCUS PHASES.

| Phases | Applications |
|---|---|
| Production | Cheetah, BTF, Spoiler Alert. |
| Distribution | Leloca, BTF, S-Cambia Cibo, ShareCity, Food Cowboy, Spoiler Alert. |
| Consumption | Leloca, Breading, BTF, Ratatouille, Polar Bear, MintScraps, WiseUpOnWaste, Menus4Mums, FitterCritters, LoveFoodHateWaste, Smart Saving, S-Cambia Cibo, ShareCity, Food Cowboy, Spoiler Alert, Resourceful Chef, Share Your Meal, Leftover Swap, MadeFood, FoodStar, CropsMobster, Foodsharing, Hello Compost, The Kroger Co., MyFoody, FameZero, LastMinuteSottoCasa, IFoodShare, SenzaSpreco, FoodCloud, Gojee. |

A. User Needs

An important observation across the surveyed applications is the user needs they meet and its extend. Some applications are primarily focused on fulfilling user needs to waste less food. The LOVE FOOD HATE WASTE and CHEETAH are examples of applications aim to reduce the overall food waste. On the other hand, other applications have added extra features to fulfill user needs and decrease the genera-tion of food waste. For instance, LELOCA, MENUS4MUMS, THE RESOURCEFUL CHEF, FOODSTAR, MYFOODY and SENZASPRECO are all applications that provide promotional coupons as extra benefits from using their services, in addition to food waste reduction. Moreover, other approaches relied on user volunteer willingness to support others in needs. For instance, BREADING, BRINGTHEFOOD, RATATOUILLE, S-CAMBIA CIBO, SHARECITY and FOOD COWBOY provide donation of surplus food as a way to motivate users reduce waste and help others.

Applications that require additional time, effort and cost are even less likely to succeed unless they can fulfill additional user needs [23]. For example, HELLO COMPOST service requires a large alteration of routine (collecting and delivering kitchen waste) or MENUS4MUMS, which necessitate reorganising areas of our life (as meals are preplanned at the beginning of the week) will be successful only if they can fulfill additional user needs, such as receiving free food (in HELLO COMPOST) or saving time (in MENUS4MUMS).

B. User Engagement

Majority of people are for wasting less food, but social pressure is often a barrier to user activity towards food waste reduction. To illustrate, people overbuy fresh fruits out of desire to be healthy. Based on the findings, few applications have considered using techniques to encourage users be more attentive when going food shopping and engage them in monitoring their waste or donation. For example, POLAR BEAR, MINTSCRAPS, FITTER CRITTERS, and HELLO COMPOST have used game design techniques to motivate their users engage with the application in performing certain activities, whereas systems, such as SHARE YOUR MEAL, SPOILER ALERT, WISE UP ON WASTE, MENUS4MUMS, LOVE FOOD HATE WASTE and SMART SAVING have used self-management and quantification, in addition to gaming techniques. It is worth mentioning that applications that only rely on the desire to reduce food waste are less likely to succeed with large mainstream audience.

C. User Awareness and Knowledge

The survey revealed few applications that focus on increasing user awareness on food waste burden. For example,

POLAR BEAR increases user's knowledge and awareness of energy saving in general, whereas FITTER CRITTERS and HELLO COMPOST promote healthy diet in the community and increase their knowledge about the importance of fruit and vegetable consumption.

D. User Attitude and Behaviour Change

The majority of applications in the survey were focused on household behaviour (shopping, eating and food preparation habits) and its influence on food waste generation, mostly on the consumption phase. Although almost all applications aim to change user behaviour towards a more responsible waste management, very few have considered taking actions in this regard. The applications didn't employ techniques to change user behaviour and increase their awareness of their food waste attitude, or even increase their engagement with the application and be more active in terms of food donation. Food waste management applications have to con-sider behavioural theories to motivate food waste reduction. Some applications (POLAR BEAR, MINTSCRAPS, FITTER CRITTERS, and HELLO COMPOST) have used simple game elements to increase user engagement, but non have consid-ered sustaining the engagement in the long-term.

E. Preventive Measure

The majority of applications covered "avoidable" food waste, like products that are still fit for human consumption the time of discarding or that would have been edible if they had been eaten in time. Few applications covered "un-avoidable" food waste, such as products that are not suited for human consumption in accordance with today's food standards (e.g., vegetable peelings, bones, egg shells). To illustrate, HELLO COMPOST focuses on "unavoidable" food waste, however, it allows users to redeem credits for fresh produce. Figure-1 illustrates the food waste management hierarchy and classifies the surveyed applications according to their focus area. For simplicity, we divide the hierarchy structure into two sections, namely waste prevention and recycling & treatment.

F. User Adaption

Surveyed applications have provided various function-alities, however, few have considered user adoption and continues use of the system. Almost no data about the average number of users involvement and application usage. Moreover, no evidence on how the application intends to increase user participation. It is important to increase user adoption and engagement with the application, since otherwise the application will be abandoned in the short-term. Focusing on adoption from the start is a hard task to achieve. Although the majority of the systems focus on user engagement, however limited number of applications (Mo-bile or Web) have focused on long-term user adoption [2]. There is a need for functionalities to boost adoption through a friendly engaging environment. For instance, gamification and game-based techniques could be an incredible way to achieve this. Using such technique helps to recognise active users and encourage others to increase their participation and move up in ranking. Gamification can reward users based on their performance. The reward system is a way to get everyone moving in the same direction. Human in nature avoid change, want to feel appreciated, get rewarded, and feel part of a community. With the rewarding or point-based gamification system, or even coupons, users will earn real prizes based on the system objective. In this way, gamification could be added as extra layer on top of food waste management systems to create an engaging environment among users. Hence, users compete for bonuses and the application will be entirely active.

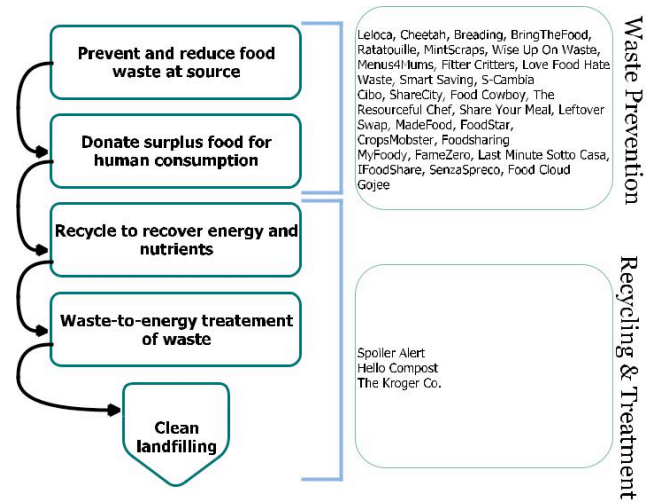

Figure 1. Waste Management Hierarchy and Applications.

A simple use case of gamification technique in food waste management applications could be as follows: imagine track-ing users donations and charities he/she contacted to help, and how much is the quantity provided. Then, the system could reward the most active user and make this visible to others and how the user has achieved the reward. On retailer or producer level, besides the reward and donation, the shop could get a badge representing the shop as a donation supporter. The badge has to reflect the shop activity and be fun, yet meaningful.

In the same way, persuasive techniques are used to motivate users towards a donation friendly behaviour. One example may be introducing simplicity and ease of use in the system. The system should provide simple user interaction and tailor the approach to reflect the way users work with the system and not the other way around. It is a bad design approach to clutter user interfaces with irrelevant info, instead give the user freedom to decide how to donate or prevent food waste.

## VI. Discussion

The majority of applications focus on food donation process, recovering food from donors (food manufacturers, distributors, retailers or individuals) and redistributing to organisations and social services supporting the most deprived. Setting up measurements for both the effectiveness of reducing waste and scaling of user uptake towards food waste should be properly assessed. There is a need for an interactive approach to design processes that avoid relying on a desire to waste less food as a primary motivation, avoid adding the pressure which currently impede food waste reduction, and rather aim to design the food waste management application to have additional motivations, beyond reduction in food waste, so that they are genuinely useful and desirable.

Food waste management applications should consider a goal beyond just food waste reduction to boost user participation. Additionally, application designers have to consider the integration of behavioural change techniques, such as persuasive technology and focus on user interaction design by making the interaction as simple as possible. Social networks could be considered to engage users in the activity. For example, the use of Facebook or Instagram to create an environment where users could contribute to food waste reduction and donation. Future developments should consider the integra-tion of a gamified layer to a core activity to achieve user motivation and long-term engagement with the application.

## VII. Conclusion

Food wastage has a significant economic, environmental and social impact. The magnitude and complexity of this problem has been tackled by several studies. This survey analysed the applications area, techniques, phases and their position in the food waste management hierarchy. Based on our findings, there is a great discrepancies between user focus and the focus of food waste management applications. Moreover, there has been little research on user behaviour and experience with food waste management applications, especially on user engagement with the application. In addition, to our knowledge the vast majority of applications have no open APIs to use for data analyses.

This observation indicates the need to incorporate techniques that move user focus beyond food waste management and create a sense of belonging and to harmonise the waste management through a set of activities. There is a need for an unobtrusive technique in food donation and by considering all age groups. Current studies have considered a generic focus on user's age and gender. For example, elderly (persons aged 65 years and above) are not adequately represented in many studies and they might be among the biggest contributors to food donation.

## References

[1] Food and A. Organization, "Global food losses and food waste - extent, causes and prevention," 2011.

[2] G. Farr-Wharton, J. H. jeong Choi, and M. Foth, "Technicolouring the fridge : reducing food waste through uses of colour-coding and cameras," in 13th International Conference on Mobile and Ubiquitous Multimedia (MUM). Melbourne, Australia: Association for Computing Machinery (ACM), 2014. [Online]. Available: http://eprints.qut.edu.au/77532/

[3] Z. Hu and J. Heidemann, "Towards geolocation of millions of IP addresses," in Proceedings of the ACM Internet Measurement Conference. Boston, MA, USA: ACM, 2012.

[4] S. Deterding, D. Dixon, R. Khaled, and L. Nacke, "From game design elements to gamefulness: Defining "gamification"," in Proceedings of the 15th Interna-tional Academic MindTrek Conference: Envisioning Future Media Environments, 2011.

[5] H. Jeff, "The rise of crowdsourcing." in Wired maga-zine 14.6 (2006): 1-4.

[6] M. Swan, "Emerging patient-driven health care models: an examination of health social networks, consumer personalized medicine and quantified self-tracking," International journal of environmental research and public health, 2009.

[7] J. Parfitt, M. Barthel, and S. Macnaughton, "Food waste within food supply chains: quantification and po-tential for change to 2050," Philosophical Transactions of the Royal Society of London B: Biological Sciences, 2010.

[8] T. Quested and P. Luzecka, "Household food and drink waste: A people focus," 2014.

[9] G. Farr-Wharton, J. H.-J. Choi, and M. Foth, "Food talks back : Exploring the role of mobile applications in reducing domestic food wastage," in OZCHI 2014 Designing Futures: The Future of Design. Syd-ney, Australia: Association for Computing Machinery (ACM), December 2014.

[10] "Cheetah," URL:http://cheetah.ujuizi.com/, accessed: 2016-08-09.

[11] "Breading," URL:http://www.breading.it/, accessed: 2016-08-09.

[12] C. A. and V. A., "Beyond food sharing: Supporting food waste reduction with icts." 2016.

[13] "Ratatouille," URL:http://www.ratatouille-app.com/, accessed: 2016-08-09.

[14] J. Bishop, "A model for understanding and influencing behaviour in virtual communities," in Proceedings of the Post-Cognitivist Psychology Conference, 2005.

[15] R. Robinson and C. Smith, "Psychosocial and demographic variables associated with consumer intention to purchase sustainably produced goods as defined by

the midwest food alliance." in Journal of Nutrition Education and Behavior, 34, 6 (2002), 316-325, 2002.

[16] J. H. Spangenberg and S. Lorek, "Environmentally sustainable household consumption: from aggregate environmental pressures to priority fields of action." in Ecological Economics, 43, 2-3 (2002), 127-140., 2002.

[17] E. Blevis, "Sustainable interaction design: Invention & disposal, renewal & reuse," in Proceedings of the SIGCHI Conference on Human Factors in Computing Systems, ser. CHI '07, 2007.

[18] D. T., B. G., M. J., and K. R., "Motivating environmentally sustainable behaviour changes with a virtual polar bear." 2008.

[19] K. C. Desouza and A. Bhagwatwar, "Technology-enabled participatory platforms for civic engagement: The case of u.s. cities," Journal of Urban Technology, 2014.

[20] "Fitter Critters," URL:http://devpost.com/software/fitter-critters, accessed: 2016-08-29.

[21] "Wise Up On Waste," URL:https://play.google.com/store/apps/details-id=uk.co.torchb2b.waste&hl=en, ac-cessed: 2016-08-29.

[22] "Menus4Mums," URL:https://www.menus4moms.com/, accessed: 2016-08-29.

[23] "Love Food Hate Wast," URL:http://www.lovefoodhatewaste.com/, accessed: 2016-08-09.

[24] "Food: Too Good to Waste Implementation Guide and Toolkit," EPA US Environmental Protection Agency, accessed: 2016-08-09. [Online]. Available: URL:https://www.epa.gov/sustainable-management-food/food-too-good-waste-implementation-guide-and-toolkit

[25] "S-Cambia Cibo," sustainable city based food sharing, accessed: 2016-08-09. [Online]. Available: URL:http: //www.scambiacibo.it/

[26] "ShareCity," URL:http://sharecity.ie/, accessed: 2016-08-09.

[27] "Food Cowboy," URL:http://www.foodcowboy.com/, accessed: 2016-08-09.

[28] "Spoiler Alert: New app designed to reduce food waste," MIT Sloan School of Management, URL:http://mitsloan.mit.edu/newsroom/articles/new-app-designed-to-reduce-food-waste, accessed: 2016-08-09.

[29] "Survey of Existing Consumer Products and Services which Reduce Food Waste," URL:http://www.shiftdesign.org.uk/content/uploads/2014/09/shiftFood-Waste survey.pdf, accessed: 2016-08-29.

[30] "Share Your Meal," URL:https://www.shareyourmeal.net/, accessed: 2016-08-29.

[31] "Leftover Swap," URL:https://play.google.com/store/apps/details-id=com.greasedwatermelon.leftoverswap&hl=en, accessed: 2016-08-29.

[32] "MadeFood,"URL:https://itunes.apple.com/us/app/made-food/id930465862-mt=8&ign-mpt=uo3D4, accessed: 2016-08-29.

[33] "CropsMobster," URL:http://sfbay.cropmobster.com/, accessed: 2016-08-29.

[34] "Foodsharing," URL:https://foodsharing.de/, accessed: 2016-08-29.

[35] "Hello Compost," URL:http://hellocompost.com/, accessed: 2016-08-29.

[36] "MyFoody," URL:https://www.myfoody.it/, accessed: 2016-08-29.

[37] "Last Minute Sotto Casa," URL:http://www.lastminutesottocasa.it/, accessed: 2016-08-29.

[38] "IFoodShare," URL:http://ifoodshare.org/, accessed: 2016-08-29.

[39] "SenzaSpreco," URL:http://info.senza-spreco.it/en/, accessed: 2016-08-29.

[40] "Food Cloud," URL:http://foodcloud.net/, accessed: 2016-08-29.